\documentclass[10pt]{article}
\usepackage{amssymb,amsmath,amsthm,mathrsfs}

\newcommand{\ie}{\emph{i.e.}}
\newcommand{\eg}{\emph{e.g.}}
\newcommand{\cf}{\emph{cf}}

\newcommand{\Real}{\mathbb{R}}
\newcommand{\pd}{\partial}
\newcommand{\Nat}{\mathbb{N}}
\newcommand{\sii}{L^2}
\newcommand{\sinf}{L^\infty}
\newcommand{\Smooth}{C}
\newcommand{\Hilbert}{\mathcal{H}}
\newcommand{\Dom}{D}
\newcommand{\diag}{\mathrm{diag}}
\newcommand{\supp}{\mathrm{supp}}

\newcommand{\spec}{\mathrm{spec}\;\!}
\newcommand{\eps}{\varepsilon}
\newcommand{\curv}{\mathcal{K}}
\newcommand{\rot}{\mathcal{R}}

\newcommand{\tube}{\mathcal{L}}
\newcommand{\vol}{\mathrm{vol}}
\newcommand{\transef}{{\mathcal{J}_1}}

\newcommand{\ds}{\partial_{\tau}}
\newtheorem{Lemma}{Lemma}
\newtheorem{Theorem}{Theorem}
\newtheorem{Proposition}{Proposition}
\newtheorem{Definition}{Definition}
\theoremstyle{definition}
\newtheorem{Remark}{Remark}
\newtheorem{ass}{Assumption}

\begin{document}
%
\title{\textbf{\Large
A Hardy inequality in twisted waveguides
}}
\author{
T.~Ekholm$^1$, \
H.~Kova\v{r}{\'\i}k$^{2,3}$
\ and \
D.~Krej\v{c}i\v{r}\'{\i}k$^3$
}
\date{
\footnotesize
\begin{quote}
\emph{
\begin{itemize}
\item[$^1$]
Department of Mathematics,
Royal Institute of Technology, \\
100\,44 Stockholm, Sweden
\item[$^2$]
Institute for Analysis, Dynamics
and Modeling, Faculty \\ of Mathematics and
Physics, Stuttgart University,
PF 80 11 40, \\ D-70569  Stuttgart, Germany
\item[$^3$]
Department of Theoretical Physics,
Nuclear Physics Institute, \\
Academy of Sciences,
250\,68 \v{R}e\v{z} near Prague, Czech Republic
\item[\emph{E-mails:}]
tomase@math.kth.se (T.E.) \\
kovarik@mathematik.uni-stuttgart.de (H.K.) \\
krejcirik@ujf.cas.cz (D.K.)
\end{itemize}
}
\end{quote}
15 December 2005
}
\maketitle
%
%
\begin{abstract}
\noindent
We show that twisting of an infinite straight
three-dimensional tube
with non-circular cross-section
gives rise to a Hardy-type inequality
for the associated Dirichlet Laplacian.
As an application we prove certain stability of the spectrum
of the Dirichlet Laplacian in locally and mildly bent tubes.
Namely, it is known that any local bending,
no matter how small, generates eigenvalues
below the essential spectrum of the Laplacian
in the tubes with arbitrary cross-sections rotated
along a reference curve in an appropriate way.
In the present paper we show
that for any other rotation some critical strength of the bending
is needed in order to induce a non-empty discrete spectrum.
\end{abstract}
%
%
%

\newpage
\section{Introduction}\label{Sec.Intro}
%
The Dirichlet Laplacian in infinite tubular domains
has been intensively studied as a model
for the Hamiltonian of a non-relativistic particle
in \emph{quantum waveguides}; we refer to
\cite{DE,LCM,Hurt} for the physical background and references.
Among a variety of results established so far,
let us point out the papers \cite{ES,GJ,RB,DE,KKriz,ChDFK}
where the existence of \emph{bound states}
generated by a local \emph{bending} of a straight waveguide is proved.
This is an interesting phenomenon for several reasons.
From the physical point of view,
one deals with a geometrically induced effect of purely quantum origin,
with important consequences for the transport in curved \emph{nanostructures}.
Mathematically, the tubes represent a class of \emph{quasi-cylindrical} domains
for which the spectral results of this type are non-trivial.

More specifically, it has been proved
in the references mentioned above that
the Dirichlet Laplacian in non-self-intersecting
tubular neighborhoods of the form
\begin{equation}\label{neighbourhood}
  \{ x\in\Real^d \ |\ \mathrm{dist}(x,\Gamma) < a \}
  \,, \qquad
  d \geq 2
  \,,
\end{equation}
where~$a$ is a positive number
and~$\Gamma$ is an infinite curve
of non-zero first curvature vanishing at infinity,
always possesses discrete eigenvalues.
On the other hand, the essential spectrum coincides as a set
with the spectrum of the straight tube of radius~$a$.
In other words, the spectrum of the Laplacian
is unstable under bending. The bound states may be generated also by other
local deformations of straight waveguides,
\eg, by a adding a ``bump'' \cite{BGRS,BEGK,EK}.

On the other hand, the first two authors of this paper
have shown recently in \cite{EK} (see also~\cite{BEK})
that a presence of an appropriate local magnetic field
in a 2-dimensional waveguide
leads to the existence of a Hardy-type inequality
for the corresponding Hamiltonian.
Consequently, the spectrum of the magnetic Schr\"odinger operator
becomes stable as a set against sufficiently weak perturbations of the type
considered above.

In this paper we show that
in tubes with non-circular cross-sections
the same stability effect
can be achieved by a purely geometrical deformation
which preserves the shape of the cross-section:
\emph{twisting}.
We restrict to $d=3$ and replace the definition~(\ref{neighbourhood})
by a tube obtained by translating an arbitrary cross-section
along a reference curve~$\Gamma$
according to a smooth \emph{moving frame} of~$\Gamma$
(\ie~the triad of a tangent and two normal vectors
perpendicular to each other).
We say that the tube is \emph{twisted}
provided (i)~the cross-section is not rotationally symmetric
(\cf~(\ref{symmetry}) below)
and (ii)~the projection of the derivative of one normal vector
of the moving frame to the other one is not zero.
The second condition can be expressed solely in terms
of the difference between the second curvature (also called torsion) of~$\Gamma$
and the derivative of the angle between the normal vectors
of the chosen moving frame and a Frenet frame of~$\Gamma$
(\cf~(\ref{noTang}) below);
the latter determines certain rotations of the cross-section
along the curve.
That is, twisting and bending may be viewed
as two independent deformations of a straight tube.
In order to describe the main results of the paper,
we distinguish two particular types of twisting.

First, when $\Gamma$ is a straight line,
then of course the curvatures are zero
and the twisting comes only from
rotations of a non-circular cross-section along the line.
In this situation, we establish Theorem~\ref{Thm.Hardy}
containing a Hardy-type inequality
for the Dirichlet Laplacian in a straight locally twisted tube.
Roughly speaking, this tells us that a local twisting stabilizes
the transport in straight tubes with non-circular cross-sections.

Second, when $\Gamma$ is curved,
the torsion is in general non-zero
and we show that it plays the same role
as the twisting due to the rotations
of a non-circular cross-section in the twisted straight case.
More specifically, we use Theorem~\ref{Thm.Hardy}
to establish Theorem~\ref{Thm.main}
saying that the spectrum of
the Dirichlet Laplacian in a twisted,
mildly and locally bent tubes coincides
with the spectrum of a straight tube,
which is purely essential.
This fact has important consequences.
For it has been proved in~\cite{ChDFK}
that any non-trivial curvature vanishing at infinity
generates eigenvalues below the essential spectrum,
provided the cross-section is translated along~$\Gamma$
according to the so-called \emph{Tang frame}
(\cf~(\ref{Tang}) below).
We also refer to~\cite{Grushin2}
for analogous results in mildly curved tubes.
But the choice of the Tang frame for the moving frame
giving rise to the tube means that
the rotation of the cross-section compensates the torsion.
Our Theorem~\ref{Thm.main} shows that this
special rotation is the only possible choice
for which the discrete eigenvalues
appear for any non-zero curvature of~$\Gamma$;
any other rotation of the cross-section
will eliminate the discrete eigenvalues
if the curvature is not strong enough.
In the curved case, we also establish Theorem~\ref{Thm.main.bis}
extending the result of Theorem~\ref{Thm.main}
to the case when also the torsion is mild.

After writing this paper,
we discovered that Grushin has in~\cite{Grushin2}
a result similar to our Theorem~\ref{Thm.main}.
Namely, using a perturbation technique
developed in~\cite{Grushin1},
he proves that there are no discrete eigenvalues
in tubes which are simultaneously
mildly curved and mildly twisted.
We would like to stress that,
apart from the different method we use,
the importance of our results lies in the fact
that the non-existence of discrete spectrum
follows as a consequence of a stronger property:
the Hardy-type inequality of Theorem~\ref{Thm.Hardy}.

The organization of the paper is as follows.
In the following Section~\ref{Sec.Results},
we present our main results; namely,
the Hardy-type inequality
(Theorem~\ref{Thm.Hardy})
and the stability result concerning the spectrum
in twisted mildly bent tubes
(Theorems~\ref{Thm.main} and~\ref{Thm.main.bis}).
The Hardy-type inequality and its local version
(Theorems~\ref{Thm.Hardy} and~\ref{hardy}, respectively)
are proved in Section~\ref{Sec.Hardy}.
In order to deal with
the Laplacian in a twisted bent tube, we have to develop
certain geometric preliminaries;
this is done in Section~\ref{Sec.tubes}.
Theorems~\ref{Thm.main} and~\ref{Thm.main.bis} are
proved at the end of Section~\ref{Sec.tubes}.
In the Appendix, we state a sufficient condition which guarantees
that a twisted bent tube does not intersect.

The summation convention is adopted throughout the paper and,
if not otherwise stated, the range of Latin and Greek indices
is assumed to be $1,2,3$ and $2,3$, respectively.
The indices~$\theta$ and~$\tau$
are reserved for a function and a vector, respectively,
and are excluded from the summation convention.
If~$U$ is an open set, we denote by~$-\Delta_D^U$
the Dirichlet Laplacian in~$U$,
\ie\ the self-adjoint operator associated in~$\sii(U)$
with the quadratic form~$Q_D^U$ defined by
$Q_D^U[\psi]:=\int_U |\nabla\psi|^2$,
$\psi\in\Dom(Q_D^U):=\Hilbert_0^1(U)$.

\section{Main results}\label{Sec.Results}
%
\subsection{Twisted bent tubes}\label{Sec.Pres.curved}
%
The tubes we consider in the present paper
are determined by a \emph{reference curve}~$\Gamma$,
a \emph{cross-section}~$\omega$
and an \emph{angle} function~$\theta$
determining a moving frame of~$\Gamma$.
We restrict ourselves to curves
characterized by their curvature functions.

Let~$\kappa_1$ and~$\kappa_2$ be $\Smooth^1$-smooth functions
on~$\Real$ satisfying
\begin{equation}\label{curvatures}
  \kappa_1 > 0
  \quad \mbox{on} \quad I
  \qquad\mbox{and}\qquad
  \kappa_1, \;\! \kappa_2 = 0
  \quad \mbox{on} \quad \Real \setminus I
  \,,
\end{equation}
where~$I$ is some fixed bounded open interval.
Then there exists a unit-speed $\Smooth^3$-smooth curve
$
  \Gamma: \Real \to \Real^3
$
whose first and second curvature functions
are~$\kappa_1$ and~$\kappa_2$, respectively
(\cf~\cite[Sec.~1.3]{Kli}).
Moreover, $\Gamma$~is uniquely determined
up to congruent transformations
and the restriction $\Gamma \upharpoonright I$
possesses a uniquely determined $\Smooth^2$-smooth
distinguished Frenet frame $\{e_1,e_2,e_3\}$.
Since the complement of $\Gamma \upharpoonright I$
is formed by two straight semi-infinite lines,
we can extend the triad $\{e_1,e_2,e_3\}$
to a $\Smooth^2$-smooth Frenet frame of~$\Gamma$.
The components~$e_1$, $e_2$ and~$e_3$
are the tangent, normal and binormal vectors of~$\Gamma$,
respectively, and~$\kappa_2$
is sometimes called the torsion of~$\Gamma$.

Given a $\Smooth_0^1$-smooth function~$\theta$ on~$\Real$,
we define the matrix valued function
\begin{equation}\label{special.rotation}
  (\rot_{\mu\nu}^\theta) =
  \begin{pmatrix}
  \cos \theta & -\sin \theta \\
  \sin \theta & \cos \theta \\
  \end{pmatrix}
  .
\end{equation}
Then the triad
$
  \{e_1,\rot_{2\nu}^\theta e_\nu,\rot_{3\nu}^\theta e_\nu\}
$
defines a $\Smooth^1$-smooth moving frame of~$\Gamma$
having normal vectors rotated by the angle~$\theta(s)$
with respect to the Frenet frame at $s\in\Real$.
Later on, a stronger regularity of~$\theta$
will be required, namely,
\begin{equation}\label{theta.regularity}
  \ddot{\theta} \in \sinf(\Real)
  \,.
\end{equation}

Let~$\omega$ be a bounded open connected set in~$\Real^2$
and introduce the quantity
\begin{equation}\label{distance}
  a := \sup_{t\in\omega} |t| \,.
\end{equation}
We assume that~$\omega$ is not rotationally invariant
with respect to the origin, \ie,
\begin{equation}\label{symmetry}
  \exists \alpha \in (0,2\pi)\,, \qquad
  \big\{
  \big( t_\mu \;\! \rot_{\mu 2}^{\alpha},
  t_\mu \;\! \rot_{\mu 3}^{\alpha} \big)
  \ |\
  (t_2,t_3)\in\omega \big\}
  \not= \omega \,.
\end{equation}

We define a \emph{twisted bent tube}~$\Omega$
about~$\Gamma$ as the image
\begin{equation}\label{tube.image}
  \Omega
  := \tube(\Real\times\omega)
  \,,
\end{equation}
where~$\tube$
is the mapping from~$\Real\times\omega$ to~$\Real^3$
defined by
\begin{equation}\label{tube}
  \tube(s,t)
  := \Gamma(s) + t_\mu \, \rot_{\mu\nu}^\theta(s) \, e_\nu(s)
  \,.
\end{equation}
We make the natural hypotheses that
\begin{equation}\label{Ass.basic}
  a \, \|\kappa_1\|_\infty < 1
  \qquad\mbox{and}\qquad
  \tube \mbox{ is injective}
  \,,
\end{equation}
so that~$\Omega$ has indeed the geometrical meaning
of a non-self-intersecting tube;
sufficient conditions ensuring the injectivity of~$\tube$
are derived in the Appendix.

Our object of interest is
the Dirichlet Laplacian in the tube,
$-\Delta_D^\Omega$.
In the simplest case when
the tube is straight (\ie~$I=\varnothing$)
and the cross-section~$\omega$
is not rotated with respect to a Frenet frame
of the reference straight line (\ie~$\dot{\theta}=0$),
it is easy to locate the spectrum:
\begin{equation}\label{spectrum.straight}
  \spec(-\Delta_D^{\Real\times\omega})
  = [E_1,\infty) \,,
\end{equation}
where~$E_1$ is the lowest eigenvalue of the Dirichlet Laplacian in~$\omega$.

Sufficient conditions for the existence
of a discrete spectrum of $-\Delta_D^\Omega$
were recently obtained in~\cite{ChDFK,Grushin2}.
In particular, it is known from~\cite{ChDFK}
that if the cross-section~$\omega$ is rotated appropriately,
namely in such a way that
\begin{equation}\label{Tang}
  \dot{\theta} = \kappa_2
  \, ,
\end{equation}
then
any non-trivial bending (\ie~$I\not=\varnothing$)
generates eigenvalues below~$E_1$,
while the essential spectrum is unchanged.

As one of the main results of the present paper we show that
condition (\ref{Tang}) is necessary for the existence of discrete spectrum in
\emph{mildly} bent tubes with non-circular cross-sections:
\begin{Theorem}\label{Thm.main}
Given $\Smooth_0^1$-curvature functions~(\ref{curvatures}),
a bounded open connected set $\omega\subset\Real^2$ satisfying
non-symmetricity condition~(\ref{symmetry})
and a $\Smooth_0^1$-smooth angle function~$\theta$
satisfying~(\ref{theta.regularity}),
let~$\Omega$ be the tube as above satisfying~(\ref{Ass.basic}).
If
\begin{equation}\label{noTang}
  \kappa_2-\dot{\theta}\not=0
  \,,
\end{equation}
then there exists a positive number~$\eps$ such that
\begin{equation*}
  \|\kappa_1\|_\infty + \|\dot{\kappa}_1\|_\infty
  \leq \eps
  \quad \Longrightarrow \quad
  \spec(-\Delta_D^{\Omega})
  = [E_1,\infty) \, .
\end{equation*}
Here $\eps$ depends on $\kappa_2$, $\dot{\theta}$ and~$\omega$.
\end{Theorem}

An explicit lower bound for the constant~$\eps$
is given by the estimates made in Section~\ref{Sec.Proof}
when proving Theorem~\ref{Thm.main};
we also refer to Proposition~\ref{Prop.injectivity} in the Appendix
for a sufficient conditions ensuring the validity of~(\ref{Ass.basic}).

Theorem~\ref{Thm.main} tells us that twisting,
induced either by torsion or by a rotation different from~(\ref{Tang}),
acts against the attractive interaction induced by bending.
Its proof is based on a Hardy-type inequality in straight tubes
presented in the following Section~\ref{Sec.Pres.straight}.
The latter provides other variants of Theorem~\ref{Thm.main},
\eg, in the situation when also the torsion is mild:
\begin{Theorem}\label{Thm.main.bis}
Under the hypotheses of Theorem~\ref{Thm.main},
with~(\ref{noTang}) being replaced by
\begin{equation}\label{noTang.bis}
  \dot{\theta}\not=0
  \,,
\end{equation}
there exists a positive number~$\eps$ such that
\begin{equation*}
  \|\kappa_1\|_\infty + \|\dot{\kappa}_1\|_\infty
  + \|\kappa_2\|_\infty
  \leq \eps
  \quad \Longrightarrow \quad
  \spec(-\Delta_D^{\Omega})
  = [E_1,\infty) \, .
\end{equation*}
Here $\eps$ depends on $\dot{\theta},\, \omega$ and~$I$.
\end{Theorem}

We refer the reader to Section~\ref{Sec.end}
for more comments on Theorems~\ref{Thm.main} and~\ref{Thm.main.bis}.

\subsection{Twisted straight tubes}\label{Sec.Pres.straight}
%
The proof of Theorems~\ref{Thm.main} and~\ref{Thm.main.bis}
is based on the fact that a twist of a straight tube
leads to a Hardy-type inequality
for the corresponding Dirichlet Laplacian.
This is the central idea of the present paper
which is of independent interest.

By the \emph{straight tube}
we mean the product set $\Real\times\omega$.
To any radial vector $t\equiv(t_2,t_3)\in\Real^2$,
we associate the normal vector
$
  \tau(t) := (t_3,-t_2)
$,
introduce the angular-derivative operator
\begin{equation}\label{angular}
  \partial_\tau := t_3 \, \partial_2 -t_2 \, \partial_3
\end{equation}
and use the same symbol for the differential expression
$1\otimes\partial_\tau$ on $\Real \times \omega$.

Given a bounded function $\sigma:\Real\to\Real$,
we denote by the same letter the function $\sigma \otimes 1$
on $\Real \times \omega$ and consider the self-adjoint operator~$L_\sigma$
associated on $\sii(\Real\times\omega)$
with the Dirichlet quadratic form
\begin{equation}\label{form.twist}
  l_\sigma[\psi] :=
  \|\partial_1\psi-\sigma\,\partial_\tau\psi\|^2
  + \|\partial_2\psi\|^2 + \|\partial_3\psi\|^2
  \,,
\end{equation}
with
$
  \psi\in\Dom(l_\sigma) := \Hilbert_0^1(\Real\times\omega)
$,
where $\|\cdot\|$ denotes the norm in $\sii(\Real\times\omega)$.

The connection between~$L_\sigma$ and a \emph{twisted straight tube}
is based on the fact that for $\sigma=\dot{\theta}$,
$L_\sigma$ is unitarily equivalent to the Dirichlet Laplacian
acting in a tube given by~(\ref{tube.image})
for the choice $\Gamma(s)=(s,0,0)$,
after passing to the coordinates determined by~(\ref{tube}).
This can be verified by a straightforward calculation.

If $\sigma=0$,
$L_0$ is just the Dirichlet Laplacian in $\Real\times\omega$,
its spectrum is given by~(\ref{spectrum.straight})
and there is no Hardy inequality associated with
the shifted operator $L_0-E_1$.
The latter means that given any multiplication operator~$V$
generated by a non-zero, non-positive function
from $C_0^\infty (\Real \times \omega)$,
the operator $L_0 - E_1 + V$ has a negative eigenvalue.
This is also true for non-trivial~$\sigma$
in the case of circular~$\omega$
centered in the origin of~$\Real^2$,
since then the angular-derivative term in~(\ref{form.twist})
vanishes for the test functions of the form $\varphi\otimes\transef$
on $\Real\times\omega$,
where~$\transef$ is an eigenfunction of the Dirichlet Laplacian
corresponding to~$E_1$.
However, in all other situations there is always a Hardy-type
inequality:
\begin{Theorem} \label{Thm.Hardy}
Let~$\omega$ be a bounded open connected subset of~$\Real^2$
satisfying the non-symmetricity condition~(\ref{symmetry}).
Let~$\sigma$ be a compactly supported continuous function
with bounded derivatives and suppose that~$\sigma$
is not identically zero.
Then, for all $\psi \in \Hilbert^1_0(\Real\times\omega)$
and any~$s_0$ such that $\sigma(s_0) \neq 0$ we have
\begin{equation}\label{Hardy}
  l_\sigma[\psi]
  - E_1 \;\! \|\psi\|^2
  \ \geq \
  c_h \int_{\Real\times\omega}
  \frac{|\psi(s,t)|^2} {1 + (s-s_0)^2} \ ds \, dt
  \,,
\end{equation}
where~$c_h$ is a positive constant independent of~$\psi$
but depending on~$s_0$, $\sigma$ and~$\omega$.
\end{Theorem}

It is possible to find an explicit lower bound
for the constant~$c_h$; we give an estimate in~(\ref{c_h}).

The assumption that~$\sigma$ has a compact support
ensures that the essential spectrum of~$L_\sigma$
coincides with~(\ref{spectrum.straight}).
As a consequence of the Hardy-type inequality~(\ref{Hardy}),
we get that the presence of a non-trivial~$\sigma$ in~(\ref{form.twist})
represents a repulsive interaction
in the sense that there is no other spectrum
for all small potential-type perturbations
having $\mathcal{O}(s^{-2})$ decay at infinity.

As explained above,
the special choice $\sigma=\dot{\theta}$ leads to
a direct geometric interpretation of~$L_\sigma$
in connection with the twisted straight tubes.
As another application of Theorem~\ref{Thm.Hardy},
we shall apply it to the twisted bent tubes, namely,
with the choice $\sigma=\kappa_2-\dot\theta$
to prove Theorem~\ref{Thm.main}
and with $\sigma=\dot{\theta}$ to prove Theorem~\ref{Thm.main.bis}
(\cf~Section~\ref{Sec.Proof}).

\section{Hardy inequality for twisted straight tubes}\label{Sec.Hardy}
%
It follows immediately from~(\ref{form.twist})
and the inequality
\begin{equation}\label{Poincare}
  \|\nabla\varphi\|_{\sii(\omega)}^2
  \ \geq \
  E_1 \;\! \|\varphi\|_{\sii(\omega)}^2
  \,, \qquad
  \forall \varphi\in\Hilbert_0^1(\omega)
  \,,
\end{equation}
that the spectrum of~$L_\sigma$ does not start below~$E_1$.
In this section, we establish the stronger result
contained in Theorem~\ref{Thm.Hardy} in two steps.
After certain preliminaries,
we first derive a ``local" Hardy inequality (Theorem~\ref{hardy}).
Then the local result is ``smeared out''
by means of a classical one-dimensional Hardy inequality.

\subsection{Preliminaries}
%
\begin{Definition}\label{Def.lambda}
To any $\omega\in\Real^2$,
we associate the number
$$
  \lambda := \inf
  \frac{\|\nabla\varphi\|_{\sii(\omega)}^2
  - E_1 \;\! \|\varphi\|_{\sii(\omega)}^2
  + \big\| \ds \varphi \big\|_{\sii(\omega)}^2}
  {\|\varphi\|_{\sii(\omega)}^2}
  \,,
$$
where the infimum is taken over all non-zero
functions from $\Hilbert_0^1(\omega)$.
\end{Definition}
\noindent
It is clear from~(\ref{Poincare})
that~$\lambda$ is a non-negative quantity.
Our Hardy inequality is based on the fact
that~$\lambda$ is always positive
for non-circular cross-sections.
\begin{Lemma} \label{crucial}
If~$\omega$ satisfies~(\ref{symmetry}),
then $\lambda>0$.
\end{Lemma}
\begin{proof}
The quadratic form~$b$ defined on $\sii(\omega)$ by
$$
  b[\varphi] :=
  \|\nabla\varphi\|_{\sii(\omega)}^2
  - E_1 \;\! \|\varphi\|_{\sii(\omega)}^2
  + \big\|
  \ds \varphi
  \big\|_{\sii(\omega)}^2
  \,, \quad
  \varphi\in\Dom(b) := \Hilbert_0^1(\omega)
  \,,
$$
is non-negative (\cf~(\ref{Poincare})),
densely defined and closed;
the last two statements follow from the boundedness of~$\tau$
and from the fact that they hold true for the quadratic form defining
the Dirichlet Laplacian in~$\omega$.
Consequently, $b$~gives rise to a self-adjoint operator~$B$.
Moreover, since $B \geq -\Delta_D^\omega-E_1$,
and the spectrum of~$-\Delta_D^\omega$ is purely discrete,
the minimax principle implies that~$B$
has a purely discrete spectrum, too.
$\lambda$ is clearly the lowest eigenvalue of~$B$.
Assume that $\lambda=0$.
Then, firstly, the ground
state $\varphi$ of~$B$ and $-\Delta_D^\omega$ coincide,
hence $\varphi$ is analytic and positive in $\omega$;
secondly, we have
$
  \ds \varphi = 0
$.
This implies that the angular derivative of $\varphi$ is zero.
Together with our assumption on $\omega$
we can conclude that there is a
point in~$\omega$ where~$\varphi$ vanishes.
This contradicts the positivity of~$\varphi$.
\end{proof}

Next we need a specific lower bound for the spectrum
of the Schr\"odinger operator in a bounded one-dimensional interval
with Neumann boundary conditions
and a characteristic function of a subinterval as the potential.
\begin{Lemma} \label{Birman}
Let~$\Lambda$ be a bounded open interval of~$\Real$.
Then for any open subinterval $\Lambda'\subset \Lambda$
and any $f \in \Hilbert^1(\Lambda)$,
the following inequality holds:
\begin{equation*}
  \|f\|_{\sii(\Lambda)}^2
  \leq c\left(\Lambda, \Lambda'\right)
  \left(
  \|f\|_{\sii(\Lambda')}^2 + \|f'\|_{\sii(\Lambda)}^2
  \right),
\end{equation*}
where
$
  c(\Lambda, \Lambda')
  := \max \left\{ 2+ 16\,(|\Lambda|/|\Lambda'|)^{2},
  4\,|\Lambda|^2 \right\}
$.
\end{Lemma}
\begin{proof}
Without loss of generality,
we may suppose that $\Lambda':=(-b/2,b/2)$ with some positive~$b$.
Define a function~$g$ on~$\Lambda$ by
$$
  g(x) :=
  \begin{cases}
    2 \, |x| / b
  & \text{for } |x| \leq b
  \,,
  \\
  1
  & \text{otherwise}
  \,.
\end{cases}
$$
Let~$f$ be any function from $\Hilbert^1(\Lambda)$.
Then $(fg)(0)=0$ and the Cauchy-Schwarz inequality gives
\begin{equation} \label{fg}
  |f(x)g(x)|^2
  \leq |x| \int_{0}^{x}\, |(fg)'|^2
  \leq |\Lambda| \, \|(fg)'\|_{\sii(\Lambda)}^2
\end{equation}
for any $x \in \Lambda$.
Now we write $f=fg+f(1-g)$ to get
$$
  \|f\|_{\sii(\Lambda)}^2
  \leq 2\, \|fg\|_{\sii(\Lambda)}^2
  + 2\, \|f(1-g)\|_{\sii(\Lambda)}^2
  = 2\, \|fg\|_{\sii(\Lambda)}^2
  +2\, \|f\|_{\sii(\Lambda')}^2
  \,.
$$
Using the estimate (\ref{fg}) and the fact that $|g'| = 2\,
|\Lambda'|^{-1}$ on $\Lambda'$,
we obtain the statement of the lemma.
\end{proof}
%

\subsection{A local Hardy inequality}
%
Since $\sigma$ is continuous and has compact support
there are closed intervals $A_j$ such that
$$
\supp \, (\sigma)= \bigcup_{j\in K} A_j
\quad \text{and} \quad
|A_i \cap A_j| = 0 \, , \ i \neq j\, ,
$$
where $K\subseteq\Nat$ is an index set.
The main result of this subsection is the following
local type of Hardy inequality:
\begin{Theorem} \label{hardy}
Let the assumptions of Theorem~\ref{Thm.Hardy} hold. For every $j\in K$
there is a positive constant~$a_j$
depending on $\sigma \upharpoonright A_j$
such that for all $\psi\in\Hilbert_0^1(\Real\times\omega)$,
\begin{equation} \label{hardy1}
  \int_{A_j\times \omega}
  \big(
  |\partial_2\psi|^2 + |\partial_3\psi|^2
  + |\pd_1 \psi + \sigma \, \ds \psi|^2
  - E_1 \;\! |\psi|^2
  \big)
  \ \geq \ a_j \, \lambda
  \int_{A_j\times\omega} |\sigma \, \psi |^2
  \,,
\end{equation}
where~$\lambda$ is the positive constant
from Definition~\ref{Def.lambda}
depending only on the geometry of~$\omega$.
\end{Theorem}

To prove Theorem~\ref{hardy},
it will be useful to introduce the following quantities:
\begin{Definition} \label{def}
For any $M\subseteq\Real$ and $\psi\in\Hilbert_0^1(\Real\times\omega)$,
we define
\begin{equation*}
\begin{aligned}
  I^M_1 &:= \|\chi_{M}\nabla'\psi \|^2 - E_1\,\|\chi_{M}\psi\|^2
  \,, \quad
  &  I^M_3 &:=  \| \chi_{M}\sigma \, \ds \psi \|^2
  \,,
  \\
  I^M_2 &:= \|\chi_{M} \pd_1\psi \|^2
  \,,
  & I^M_{2,3} &:= - 2\,\Re\, (\pd_1\psi,\chi_{M}\sigma\,\ds\psi)
  \,,
\end{aligned}
\end{equation*}
where $\chi_{M}$ denotes the characteristic function
of the set $M \times \omega$,
$\nabla'$~denotes the gradient operator
in the ``transverse'' coordinates $(t_2,t_3)$
and $(\cdot,\cdot)$ is the inner product
generated by~$\|\cdot\|$.
\end{Definition}
\noindent
Note that~$I^M_1$ is non-negative due to~(\ref{Poincare})
and that we have
\begin{equation}\label{form.again}
  l_\sigma[\psi] - E_1 \;\! \|\psi\|^2
  =  I_1^\Real + I_2^\Real + I^{\supp(\sigma)}_3 + I^{\supp(\sigma)}_{2,3}
  \,.
\end{equation}
Let $A$ be the union of any (finite or infinite) sub-collection
of the intervals $A_j$.

The following lemma enables us to estimate the mixed term~$I_{2,3}^A$.
\begin{Lemma} \label{mixedterm}
Let the assumptions of Theorem \ref{Thm.Hardy} be satisfied.
Then for each positive numbers $\alpha$ and $\beta$,
there exists a constant $\gamma_{\alpha,\beta}$
depending also on~$\sigma\restriction A$
such that for any $\psi\in\Hilbert_0^1(\Real\times\omega)$,
\begin{equation*} 
  |I^A_{2,3}|
  \ \leq \
  \gamma_{\alpha,\beta}\, I^A_1+\alpha\, I^B_2+\beta\,  I^A_3
  \,,
\end{equation*}
where $B:=(\inf A,\, \sup A)$.
\end{Lemma}
\begin{proof}
It suffices to prove the result for real-valued functions~$\psi$
from the dense subspace $\Smooth_0^\infty(\Real\times\omega)$.
We employ the decomposition
\begin{equation} \label{factorisation}
  \psi (s,t) = \transef (t)\, \phi (s,t)
  \,, \qquad
  (s,t) \in \Real\times\omega
  \,,
\end{equation}
where $\transef$ is a positive eigenfunction of the Dirichlet Laplacian
on~$\sii(\omega)$ corresponding to~$E_1$
(we shall denote by the same symbol
the function $1\otimes\transef$ on $\Real\times\omega$),
and~$\phi$ is a real-valued function
from $\Smooth_0^\infty(\Real\times\omega)$,
actually introduced by~(\ref{factorisation}).
Then
\begin{equation*}
\begin{aligned}
  I^A_1 &= \| \chi_{A} \transef \nabla' \phi \|^2
  \,, \quad
  &  I^A_3 &= \| \chi_{A} \sigma (\transef \ds \phi + \phi \;\! \ds
  \transef )\|^2
  \,,
  \\
  I^A_2 &= \| \chi_{A} \transef \pd_1 \phi\|^2
  \,,
  & I^A_{2,3} &= - 2\,\big( \transef \pd_1\phi,
  \chi_{A} \sigma \, ( \transef \ds\phi
  + \phi \;\! \ds\transef) \big)
  \,,
\end{aligned}
\end{equation*}
where we have integrated by parts to establish
the identity for~$I^A_1$.
Using
$$
  |\sigma\, \ds \phi|^2 \leq c_1 \, |\nabla' \phi|^2
  \,, \qquad \mbox{with} \quad
  c_1 := \|\sigma \restriction A\|^2_\infty \, a^2
  \,,
$$
and applying the Cauchy-Schwarz inequality
and the Cauchy inequality with $\alpha>0$,
the first term in the sum of~$I^A_{2,3}$
can be estimated as follows:
\begin{equation}\label{first}
  \big| 2 \!\;
  \big(\transef \pd_1 \phi , \chi_{A} \sigma \, \transef\, \ds \phi
  \big)
  \big|
  \leq 2 \, \sqrt{c_1} \sqrt{I^A_1} \sqrt{I^A_2} \,
  \leq \frac{2\,c_1}{\alpha}\, I^A_1
  + \frac{\alpha}{2} \, I^A_2
  \,.
\end{equation}
In order to estimate the second term,
we first combine integrations by parts to get
$$
  \big| 2 \!\;
  \big( \transef \pd_1 \phi, \chi_{A} \sigma \, \phi \, \ds \transef
  \big)
  \big| = \big|\big( \phi, \chi_{A} \dot\sigma \;\! \transef^2 \,
  \ds \phi \big) \big|
  \,.
$$
Using
$$
  |\dot\sigma\, \ds \phi |^2 \leq c_2 \, |\nabla'\phi|^2
  \,, \qquad \mbox{with} \quad
  c_2 := \|\dot\sigma \restriction A\|^2_\infty \, a^2
  \,,
$$
and the Cauchy-Schwarz inequality, we have
$$
  \big|\big( \phi, \chi_{A} \dot\sigma \;\! \transef^2 \,
  \ds \phi \big) \big|^2 \leq c_2\, I^A_1\, \| \chi_{A} \transef \phi\|^2
  \,,
$$
Obviously, we can find an open interval $A' \subset A$
such that there exists a certain positive number~$\sigma_0$,
for which
$$
\sigma(s)\geq \sigma_0, \qquad \forall\, s\in A'\, .
$$
Lemma \ref{Birman} tells us that
\begin{eqnarray*}
\| \chi_{A} \, \transef \phi\|^2 \leq \| \chi_{B} \, \transef \phi\|^2 & \leq &
  c(B, A') \, \big(I^B_2 + \|\chi_{A'} \,
  \transef \phi\|^2 \big) \\
& \leq &  c(B, A')\,
\big (I^B_2 + \sigma_0^{-2} \|\chi_{A'} \, \sigma
  \transef \phi\|^2 \big)\, .
\end{eqnarray*}
Moreover, for each fixed value of $s\in\Real$ we have
$\sigma(s)\,\transef\, \phi(s,\cdot) \in \Hilbert_0^1(\omega)$,
and therefore we can apply Lemma \ref{crucial} to obtain
\begin{equation*}
\| \chi_{A'} \, \sigma \transef \phi\|^2 \leq
\| \chi_{A} \sigma \transef \phi\|^2 \leq \lambda^{-1} \, \big( I^A_3 +
\|\sigma\|_\infty^2 \, I^A_1\big)\, .
\end{equation*}
Writing $c_3:= c_2\, c(B, A') \lambda^{-1} \sigma_0^{-2}$,
we conclude that
\begin{eqnarray}\label{second}
  \big|\big( \phi, \chi_{A}\, \dot\sigma \;\! \transef^2\, \ds \phi \big)
  \big|^2
  &\leq& c_3 \, I^A_1\, \big( \|\sigma\|_\infty^2\, I^A_1 + \lambda\,
  \sigma_0^2\, I^B_2 + I^A_3 \big)
  \nonumber \\
  &\leq&
  \left(\tilde{\gamma}_{\alpha,\beta} \, I^A_1 +
  \frac{\alpha}{2}\, I^B_2 + \beta\, I^A_3\right )^2
\end{eqnarray}
for any $\beta > 0$ and
$
  \tilde\gamma_{\alpha,\beta} := \max \{ \sqrt{c_3}\,
  \|\sigma\|_\infty\, ,\, c_3 (2\beta)^{-1}\, ,\, c_3\, \lambda\, \sigma_0^2
  \, \alpha^{-1} \}
$.
Finally, combining~(\ref{first}) with~(\ref{second}),
the estimate for~$|I^A_{2,3}|$ follows by setting
$
  \gamma_{\alpha,\beta}
  := \tilde{\gamma}_{\alpha,\beta} + 2 \, c_1 \alpha^{-1}
$.
\end{proof}

Now we are in a position to establish Theorem~\ref{hardy}.
\begin{proof}[Proof of Theorem~\ref{hardy}]
We take $A=A_j$,
$\alpha=1$, $\beta<1$ and keep in mind that $\gamma_{1,\beta}$ in
Lemma~\ref{mixedterm} depends on $j$. We
define  $\gamma(\beta,j):=\max\{1/2,\gamma_{1,\beta}\}$.
Lemma~\ref{mixedterm} then gives
\begin{eqnarray*}
&&  \int_{A_j\times \omega} \big(|\nabla'\psi|^2 + |\pd_1 \psi + \sigma \,
  \ds \psi|^2 - E_1 \, |\psi|^2 \big)  \\
&& \qquad \qquad \geq
  \frac{1}{2} \, I^{A_j}_1
  + \left(1-\frac{1}{2\gamma(\beta,j)}\right)
  \big( I^{A_j}_2 + I^{A_j}_3 - |I^{A_j}_{2,3}| \big)
  + \frac{1-\beta}{2\gamma(\beta,j)} \, I^{A_j}_3
\end{eqnarray*}
Since
$
  I^{A_j}_2 + I^{A_j}_3 - |I^{A_j}_{2,3}| \geq 0
$,
we get from Lemma \ref{crucial} that
\begin{eqnarray*}
&& \int_{A_j\times \omega} \big(|\nabla'\psi|^2 + |\pd_1 \psi + \sigma \,
  \ds \psi|^2 - E_1 |\psi|^2 \big) \\
&& \qquad \qquad \qquad \qquad  \ \geq \ a_j\, \big(\|\sigma \restriction
  A_j\|^2_\infty \,
  I^{A_j}_1+I^{A_j}_3\big)  \ \geq \  a_j\, \lambda\,  \int_{A_j\times\omega}
  |\sigma \, \psi |^2
  \,,
\end{eqnarray*}
where
\begin{equation*}
  a_j=\frac 12\, \min\left\{\frac{1}{\|\sigma \restriction A_j\|^2_\infty}\, ,\,
  \frac{1-\beta}{\gamma(\beta,j)}\right\}\, .
\end{equation*}
\end{proof}

\begin{Remark}
Note that the Hardy weight on the
right hand side of~(\ref{hardy1}) cannot be made arbitrarily
large by increasing $\sigma$, since  the constant~$a_j$
is proportional to $\|\sigma \restriction A_j\|_{\infty}^{-2}$
if the latter is large enough. We want to point out that this degree of decay
of $a_j$ is optimal if the axes of rotation intersects $\omega$. Assume there
exists an $\alpha<2$, such that $a_j$ is proportional to $\|\sigma \restriction
A_j\|_{\infty}^{-\alpha}$ when $\|\sigma \restriction A_j\|_{\infty}\to\infty$.
Consider a test function~$\psi$ of the form
$
  \psi(s,t) := g(s) f(t)
$,
where $g\in\Hilbert^1(\Real)$ is supported inside $A_j$
and $f\in\Hilbert_0^1(\omega)$ is radially
symmetric with respect to the intersection of $\omega$ with
the axes of rotation. Then $\ds\psi = 0$ on $A_j\times\omega$
and therefore the left hand side of~(\ref{hardy1})
is for this test function independent of $\sigma$.
Take $\sigma= n\, \tilde{\sigma}$ with $\tilde{\sigma}$ being a fixed
function. The right hand side of (\ref{hardy1}) then tends to infinity as
$n\to\infty$ which contradicts the inequality.
\end{Remark}
%

\subsection{Proof of Theorem \ref{Thm.Hardy}}
%
For applications, it is convenient to replace
the Hardy inequality of Theorem~\ref{hardy}
with a compactly supported Hardy weight by a global one.
To do so, we recall the following version of
the one-dimensional Hardy inequality:
\begin{equation}\label{hardycl}
\int_{\Real}\, \frac{|v(x)|^2}{x^2} \, dx \leq 4\, \int_{\Real}\,
|v'(x)|^2\, dx
\end{equation}
for all $v\in C_0^{\infty}(\Real)$ with $v(0)=0$. Inequality (\ref{hardycl})
extends by continuity to all $v\in\Hilbert^1(\Real)$ with $v(0)=0$.

Without loss of generality we can assume that $s_0 = 0$.
Let $J = [-b,b]$, with some positive number~$b$,
be an interval where $|\sigma|>0$.
Let $\tilde{f}:\Real \to \Real$ be defined by
\begin{equation*}
  \tilde{f}(s) :=
\begin{cases}
  1
  & \mbox{for } \ |s| \geq b \,,
  \\
  |s|/b
  & \mbox{for } \ |s| < b \,,
\end{cases}
\end{equation*}
and put $f:=\tilde{f} \otimes 1$ on $\Real\times\omega$.
For any $\psi\in\Smooth_0^\infty(\Real\times\omega)$,
let us write $\psi = f\psi + (1-f)\psi$.
Applying~(\ref{hardycl}) to the function
$s \mapsto (f\psi)(s,t)$ with~$t$ fixed,
we arrive at
\begin{align*}
\int_{\Real\times\omega} \frac{|\psi(s,t)|^2} {1 + s^2} \, ds \, dt
&\leq 2 \int_{\Real\times\omega} \frac{|\tilde{f}(s)\psi(s,t)|^2} {s^2} \, ds
\, dt +
2 \int_{J\times\omega} |(1-f)\psi|^2 \\
&\leq 16 \, \|(\partial_1 f)\psi\|^2
+ 16 \, \|f \pd_1\psi\|^2 \, + 2 \, \| \chi_{J}
(1-f)\psi\|^2 \\
&\leq \left( \frac{16}{b^2} + 2\right) \| \chi_{J} \psi\|^2
+ 16\, \|\pd_1\psi\|^2 ,
\end{align*}
where $\chi_J$ denotes the characteristic function of the set $J\times\omega$.
Theorem~\ref{hardy} then implies that there exists a positive constant $c_0$
depending on $\sigma$ such that
$$
  \| \chi_{J} \psi\|^2
  \leq \big(c_0\;\!\lambda\;\!\min_J|\sigma|\big)^{-1}
  \big(Q_0[\psi] - E_1 \|\psi\|^2\big)
  \, .
$$
To estimate the second term we let $A = \supp \, (\sigma)$
and rewrite the inequality of Lemma~\ref{mixedterm}
for $\beta=1$ as
$$
  \gamma_{\alpha}^{-1} \, |I^A_{2,3}|
  \leq I^A_1 + \alpha \, \gamma_\alpha^{-1} \, I^B_2
  + \gamma_\alpha^{-1} \, I^A_3
  \,,
$$
where $\gamma_\alpha := \max\{1,\gamma_{\alpha,1}\}$
and $\alpha\in(0,1)$.
Substituting this inequality into~(\ref{form.again}),
writing
$
  I^A_{2,3} = \gamma_{\alpha}^{-1} \, I^A_{2,3}
  + (1-\gamma_{\alpha}^{-1}) \, I^A_{2,3}
$
and employing $I^A_2 + I^A_3 + I^A_{2,3}\geq 0$,
we obtain
$$
  I^B_2 = \|\chi_B\, \pd_1\psi\|^2
  \leq
  \gamma_\alpha \,(1-\alpha)^{-1} \,
  \big(Q_0[\psi] - E_1 \|\psi\|^2\big)
  \,.
$$
On the complement of $B\times\omega$ we have a trivial estimate
$$
 \|\chi_{\Real\setminus B}\, \pd_1\psi\|^2 \leq Q_0[\psi] - E_1 \|\psi\|^2\, .
$$
Summing up, the density of $\Smooth_0^\infty(\Real\times\omega)$
in $\Hilbert_0^1(\Real\times\omega)$ implies Theorem~\ref{Thm.Hardy}
with
\begin{equation}\label{c_h}
  c_h \geq
   \left[\frac{16+2\;\!b^2}{b^2\,c_0\,
  \lambda\,\min_{J} |\sigma|^2}
  + 16\, \left( \frac {\gamma_{\alpha}}{1-\alpha}+1\right)\right]^{-1}
  \,.
\end{equation}
%

\section{Twisted bent tubes}\label{Sec.tubes}
%
Here we develop a geometric background
to study the Laplacian in bent and twisted tubes,
and transform the former into a unitarily equivalent
Schr\"odinger-type operator in a straight tube.
At the end of this section,
we also perform proofs of Theorems~\ref{Thm.main} and~\ref{Thm.main.bis}
using Theorem~\ref{Thm.Hardy}.
We refer to Section~\ref{Sec.Pres.curved} for definitions
of basic geometric objects used throughout the paper.

While we are mainly interested in the curves
determined by curvature functions of type~(\ref{curvatures}),
we stress that the formulae
of Sections~\ref{Sec.metric} and~\ref{Sec.Laplacian}
are valid for arbitrary curves
(it is only important to assume the existence of
an appropriate Frenet frame for the reference curve of the tube,
\cf~\cite{ChDFK}).

\subsection{Metric tensor}\label{Sec.metric}
%
Assuming~(\ref{Ass.basic}) and using the inverse function theorem,
we see that the mapping~$\tube$ introduced in~(\ref{tube})
induces a $\Smooth^1$-smooth diffeomorphism
between the straight tube $\Real\times\omega$
and the image~$\Omega$.
This enables us to identify~$\Omega$
with the Riemannian manifold $(\Real\times\omega,G_{ij})$,
where~$(G_{ij})$ is the metric tensor induced by the embedding~$\tube$,
\ie\
$$
  G_{ij}
  := (\pd_i\tube) \cdot\, (\pd_j\tube)
  \,,
$$
with the \emph{dot} being the scalar product in~$\Real^3$.

Recall the Serret-Frenet formulae (\cf~\cite[Sec.~1.3]{Kli})
\begin{equation}\label{Frenet}
  \dot{e}_i = \curv_{ij} \, e_j \,,
  \qquad
  i \in \{1,2,3\} \,,
\end{equation}
where the matrix-valued function ~$(\curv_{ij})$
has the skew-symmetric form
\begin{equation}\label{curvature}
  (\curv_{ij}) =
\begin{pmatrix}
   0 & \kappa_1 & 0 \\
   -\kappa_1 & 0 & \kappa_2 \\
   0 & -\kappa_2 & 0
\end{pmatrix}
  .
\end{equation}
Using~(\ref{Frenet}) and the orthogonality conditions
$
  \rot_{\mu\rho}^\theta \rot_{\rho\nu}^\theta = \delta_{\mu\nu}
$,
we find
\begin{equation}\label{metric}
  (G_{ij}) =
  \begin{pmatrix}
    h^2+h_\mu h_\mu  & h_2 & h_3 \\
    h_2 & 1 & 0 \\
    h_3 & 0 & 1 \\
  \end{pmatrix}
  ,
\end{equation}
where
\begin{align*}
  h(s,t)
  &:= 1 - [t_2\cos\theta(s)+t_3\sin\theta(s)] \, \kappa_1(s) \,,
  \\
  h_2(s,t)
  &:= - t_3 \, [\kappa_2(s)-\dot{\theta}(s)] \,,
  \\
  h_3(s,t)
  &:= t_2 \, [\kappa_2(s)-\dot{\theta}(s)] \,.
\end{align*}
Furthermore,
$$
  G :=\det(G_{ij})=h^2
  \,,
$$
which defines the volume element of $(\Real\times\omega,G_{ij})$ by setting
$$
  d\vol := h(s,t)\,ds\,dt.
$$
Here and in the sequel $dt \equiv dt_2 \, dt_3$
denotes the $2$-dimensional Lebesgue measure in~$\omega$.

The metric is uniformly bounded and elliptic
in view of the first of the assumptions in~(\ref{Ass.basic});
in particular, (\ref{distance}) yields
\begin{equation}\label{1<G<1}
  0 <
  1 - a \, \|\kappa_1\|_\infty
  \ \leq \ h \ \leq \
  1 + a \, \|\kappa_1\|_\infty
  < \infty
  \,.
\end{equation}
It can be directly checked that the inverse $(G^{ij})$
of the metric tensor~(\ref{metric}) is given by
\begin{equation}\label{inverse.metric}
  (G^{ij}) = \frac{1}{h^2}
  \begin{pmatrix}
    1 & -h_2 & -h_3 \\
    -h_2 & h^2+h_2^2 & h_2 h_3 \\
    -h_3 & h_3 h_2 & h^2+h_3^2 \\
  \end{pmatrix}
  .
\end{equation}
It is worth noticing that one has the decomposition
\begin{equation}\label{positive}
  (G^{ij}) = \diag(0,1,1) + (S^{ij})
  \,,
\end{equation}
where the matrix~$(S^{ij})$ is positive semi-definite.

\subsection{The Laplacian}\label{Sec.Laplacian}
%
Recalling the diffeomorphism between
$\Real\times\omega$ and~$\Omega$ given by~$\tube$,
we identify the Hilbert space~$\sii(\Omega)$
with $\sii(\Real\times\omega,d\vol)$.
Furthermore, the Dirichlet Laplacian~$-\Delta_D^{\Omega}$
is unitarily equivalent to the self-adjoint operator~$\tilde{Q}$
associated on $\sii(\Real\times\omega,d\vol)$
with the quadratic form
\begin{equation}\label{form}
  \tilde{q}[\psi] := \int_{\Real\times\omega}
  (\overline{\pd_i\psi}) \, G^{ij} \, (\pd_j\psi) \ d\vol
  \,,
  \qquad
  \psi\in\Dom(\tilde{q})
  := \Hilbert_0^1(\Real\times\omega,d\vol)
  \,.
\end{equation}
We can write
$
  \tilde{Q}
  = -G^{-1/2} \partial_i G^{1/2} G^{ij} \partial_j
$
in the form sense,
which is a general expression for the Laplace-Beltrami operator
in a manifold equipped with a metric~$(G_{ij})$.

Now we transform~$\tilde{Q}$ into a unitarily equivalent
operator~$Q$ acting in the Hilbert space $\sii(\Real\times\omega)$,
without the additional weight~$G^{1/2}$ in the measure of integration.
This is achieved by means of the unitary operator
\begin{equation*}
  \mathcal{U}:
  \sii(\Real\times\omega,d\vol) \to \sii(\Real\times\omega) :
  \big\{ \psi \mapsto G^{1/4}\psi \big\}
  \,.
\end{equation*}
Defining
$
  Q := \mathcal{U}\,\tilde{Q}\,\mathcal{U}^{-1}
$,
it is clear that~$Q$ is the operator associated
with the quadratic form
$$
  q[\psi] := \tilde{q}[G^{-1/4}\psi]
  \,, \qquad
  \psi \in \Dom(q) := \Hilbert_0^1(\Real\times\omega)
  \,.
$$
It is straightforward to check that
\begin{equation}\label{form.bis}
  q[\psi]
  = \big(\pd_i\psi,G^{ij}\pd_j\psi\big)
  + \big(\psi,(\pd_i F) G^{ij} (\pd_j F) \,\psi\big)
  + 2 \, \Re\,\big(\pd_i\psi,G^{ij} (\pd_j F) \, \psi\big)
  \,,
\end{equation}
where
\begin{equation*}
  F := \log(G^{1/4})
  \,.
\end{equation*}
%

\subsection{Proof of Theorems~\ref{Thm.main}
and~\ref{Thm.main.bis}}\label{Sec.Proof}
%
Let us first prove Theorem~\ref{Thm.main}.
Putting $\sigma:=\kappa_2-\dot\theta$,
we observe that~$l_\sigma$ is equal to~$q$ after letting
$
  k := \|\kappa_1\|_\infty + \|\dot{\kappa}_1\|_\infty
$
equal to zero in the latter form.
Hence, the proof of Theorem~\ref{Thm.main}
reduces to a comparison of these quadratic forms
and the usage of Theorem \ref{Thm.Hardy}.
Let $(G_0^{ij})$ be the matrix~(\ref{inverse.metric})
after letting $\kappa_1=0$,
\ie\ with~$h$ being replaced by~$1$
while~$h_2$ and~$h_3$ being unchanged;
then
$
  l_\sigma[\psi]
  = (\partial_i \psi,G_0^{ij}\partial_j\psi)
$.

We strengthen the first of the hypotheses~(\ref{Ass.basic}) to
$$
  \|\kappa_1\|_\infty \leq 1/(2a)
  \,,
$$
so that we have a uniform positive lower bound to~$h$,
namely $h \geq 1/2$ due to~(\ref{1<G<1}).
It is straightforward to check
that we have on $\Real\times\omega$
the following pointwise inequalities:
\begin{align*}
  \max_{i,j\in\{1,2,3\}}
  |G^{ij}-G_0^{ij}|
  &\leq C_1 \, k \, \chi_I \,,
  \\
  \max_{i\in\{1,2,3\}}
  |\pd_i F|
  &\leq C_2 \, k \, \chi_I
  \,,
\end{align*}
where~$\chi_I$ denotes the characteristic function
of the set $I\times\omega$ and
\begin{equation*}
  C_1
  := 6 \, a \,
  \big(
  1+ a \, \|\kappa_2-\dot\theta\|_\infty
  \big)^2
  \,, \qquad
  C_2
  := 1 + a \, \big( 1 + \|\dot\theta\|_\infty \big)
  \,.
\end{equation*}
At the same time,
\begin{equation*}
  C_3^{-1} \, 1 \leq (G_0^{ij}) \leq C_3 \, 1
  \,,
\end{equation*}
in the matrix-inequality sense on $\Real\times\omega$,
where~$1$ denotes the identity matrix and
$$
  C_3
  := 1 + a \, \|\kappa_2-\dot\theta\|_{\infty} +
  a^2\,\|\kappa_2-\dot\theta\|_{\infty}^2
  \,.
$$
Consequently,
we have the following matrix inequality on $\Real\times\omega$:
\begin{equation*}
  (1-C_4\,k\,\chi_I) (G_0^{ij})
  \leq (G^{ij}) \leq
  (1+C_4\,k\,\chi_I) (G_0^{ij})
  \,,
\end{equation*}
where $C_4:=3 \;\! C_1 C_3$.
Finally, if we assume that $k \leq 1$ we have
\begin{equation}\label{F.estimate}
  |(\pd_i F)G^{ij}{(\pd_jF)}|
  \leq C_5^2\,k^2 \, \chi_I
  \,,
\end{equation}
where
$
  C_5
  := C_2 \sqrt{3\;\!C_3(1+C_4)}
$.

Let~$\psi$ be any function from $\Hilbert_0^1(\Real\times\omega)$.
First we estimate the term of indefinite sign
on the right hand side of~(\ref{form.bis}) as follows:
\begin{align*}
  2 \, \big|\Re\,\big(\pd_i\psi,G^{ij}{(\pd_jF)}\,\psi\big)\big|
  & \leq 2 \, C_5 \, k
  \left(
  \chi_I
  \big[(\overline{\partial_i\psi}) G^{ij} (\partial_j\psi)\big]^{1/2} ,
  \chi_I |\psi|
  \right)
  \\
  & \leq C_5^2\,k \, \|\chi_I\psi\|^2
  + k \, \big(\pd_i\psi, \chi_I \, G^{ij}\pd_j\psi\big)
  \,.
\end{align*}
Here the first inequality is established by applying the Cauchy-Schwarz
inequality to the inner product induced by~$G^{ij}$
and using~(\ref{F.estimate}).
The second inequality follows by the Cauchy-Schwarz inequality
in the Hilbert space $\sii(\Real\times\omega)$
and by an elementary Cauchy inequality.
Consequently,
\begin{equation}\label{inter}
  q[\psi] \geq
  \big(\pd_i\psi,(1-C_6\,k\,\chi_I)\,G_0^{ij}\pd_j\psi\big)
  - C_7\,k\, \|\chi_I\psi\|^2
  \,,
\end{equation}
where
$
  C_6 := 1 + C_4
$
and
$
  C_7 := 2 \;\! C_5^2
$.

Assume $k < C_6^{-1}$,
using the decomposition of the type~(\ref{positive})
for the matrix~$(G_0^{ij})$,
neglecting the positive contribution
coming from the corresponding matrix~$(S_0^{ij})$,
using the Fubini theorem
and applying~(\ref{Poincare}) to the function
$
  \varphi :=
  \int_{\Real} \sqrt{1-C_6\,k\,\chi_I(s)}
  \, \psi(s,\cdot) \, ds
$,
we may estimate~(\ref{inter}) as follows:
\begin{equation*}
  q[\psi] - E_1 \|\psi\|^2
  \geq (1-C_6\,k) \big(l_\sigma[\psi] - E_1 \|\psi\|^2\big)
  - (C_6 E_1+C_7)\,k\,\|\chi_I\psi\|^2
  \,.
\end{equation*}
Applying Theorem~\ref{Thm.Hardy}
to the right hand side of the previous inequality,
we have
\begin{equation*}
  q[\psi] - E_1 \|\psi\|^2
  \geq
  \int_{\Real\times\omega}
  \left(
  \frac{c_h\,(1-C_6\,k)}{1+(s-s_0)^2}
  - (C_6 E_1+C_7)\,k\,\chi_I(s)
  \right)
  |\psi(s,t)|^2 \, ds\,dt
  \,,
\end{equation*}
where~$c_h$ is the Hardy constant of Theorem~\ref{Thm.Hardy}.
This proves that the threshold of the spectrum of~$Q$
(and therefore of~$-\Delta_D^{\Omega}$)
is greater than or equal to~$E_1$
for sufficiently small~$k$.

In order to show that the whole interval $[E_1,\infty)$
belongs to the spectrum,
it is enough to construct an appropriate Weyl sequence
in the infinite straight ends of~$\Omega$.
This concludes the proof of Theorem~\ref{Thm.main}.

The proof of Theorem~\ref{Thm.main.bis} is exactly the same,
provided one chooses $\sigma:=\dot\theta$ and
$
  k := \|\kappa_1\|_\infty + \|\dot{\kappa}_1\|_\infty
  + \|\kappa_2\|_\infty
$.
Indeed, all the above estimates are valid with
$(G_0^{ij})$ being now the matrix~(\ref{inverse.metric})
after letting both $\kappa_1$ and~$\kappa_2$ equal to zero,
and with~$C_1$ and~$C_3$ being replaced by
$$
  C_1
  := 6 \, a \,
  \big(
  1+ a \, \|\kappa_2\|_\infty + a \, \|\dot\theta\|_\infty
  \big)^2
  \,, \qquad
  C_3
  := \max\big\{
  2, 1 + 2\,a^2\,\|\dot\theta\|_{\infty}^2
  \big\}
  \,,
$$
respectively.
Here~$C_1$ can be further estimated
by a constant independent of~$\kappa_2$
provided one restricts, \eg, to $\|\kappa_2\|_\infty<1/a$.

\section{Conclusions}\label{Sec.end}
%
We established Hardy-type inequalities
for twisted 3-dimensional tubes.
As an application we showed that
the discrete eigenvalues of the Dirichlet Laplacian
in mildly and locally bent tubes
can be eliminated by an appropriate twisting.
However, we would like to point out that for $\sigma=\dot\theta$,
Theorems~\ref{Thm.Hardy} and~\ref{hardy}
can be used to prove certain
stability of transport in straight twisted tubes also against other types
of perturbations. For example against a local enlargement
of the straight tube, mentioned in Introduction,
or in principle against any potential
perturbation which decays at least as $\mathcal{O}(s^{-2})$ at infinity,
where~$s$ is the longitudinal coordinate of the straight tube.
The required decay at infinity is related
to the decay of the Hardy weight in Theorem~\ref{Thm.Hardy},
and cannot be therefore improved by our method.
The quadratic decay of the Hardy weight
is determined by the classical inequality (\ref{hardycl})
and is typical for Hardy inequalities for the Laplace operator.

For straight twisted tubes,
the Hardy weight in the local inequality~(\ref{hardy1})
of Theorem~\ref{hardy}
is given in terms of the the function~$\dot\theta$
and the constant~$\lambda$.
Roughly speaking, the first tells us how fast the cross-section rotates,
while the latter ``measures'' how much the cross-section
differs from a disc.
The actual value of~$\lambda$ depends of course on the geometry of~$\omega$.

The example of bent twisted tubes is of particular interest, since
it shows the important role of the torsion. Namely, Theorem~\ref{Thm.main}
tells us that, whenever $\dot\theta\not=\kappa_2$,
the discrete eigenvalues in mildly curved tubes can be
eliminated by torsion only.
Note that Theorem~\ref{Thm.main} also provides a better lower bound to the spectrum
in mildly bent tubes than that derived in~\cite{EFK}.

Theorems~\ref{Thm.main} and~\ref{Thm.main.bis}
were proved for tubes about curves
determined by~(\ref{curvatures}).
This restriction was made in order to construct the tube uniquely
from given curvature functions by means of a uniquely determined Frenet frame.
However, Theorems~\ref{Thm.main} and~\ref{Thm.main.bis}
will also hold for more general classes of tubes, namely,
for those constructed about curves possessing
the \emph{distiguished} Frenet frame
and with curvatures decaying as $\mathcal{O}(s^{-2})$ at infinity,
where~$s$ is the arc-length parameter of the curve.

At least from the mathematical point of view,
it would be interesting to extend Theorem~\ref{Thm.main}
to higher dimensions.
Here the main difficulty is that~$\sigma$
in the form analogous to~(\ref{form.twist})
will be in general a tensor depending also
on the transverse variables~$t$.
Nevertheless, a higher dimensional analogue
of Theorem~\ref{Thm.main.bis}
is easy to derive along the same lines as in the present paper,
provided one restricts to rotations of the cross-section
just in one hyperplane.

Summing up, the twisting represents
a repulsive geometric perturbation
in the sense that it eliminates the discrete eigenvalues
in mildly curved waveguides.
Regarding the transport itself,
an interesting open question is
whether this also happens to the singular spectrum
possibly contained in the essential spectrum.

\appendix
\section{Injectivity of the tube mapping}\label{Appendix}
%
Let us conclude the paper by finding geometric conditions
which guarantee the basic hypotheses~(\ref{Ass.basic}).

The first condition of~(\ref{Ass.basic})
ensures that the mapping~$\tube$
is an immersion due to~(\ref{1<G<1}).
The second, injectivity condition requires
to impose some global hypotheses about the geometry of the curve.
Our approach is based on the following lemma:
\begin{Lemma}\label{Lem.closing}
Let~$\Gamma$ be determined
by the curvature functions~(\ref{curvatures}).
Then for every $i\in\{1,2,3\}$ and all $s_1,s_2\in\Real$,
$$
  \big|e_i(s_2)-e_i(s_1)\big|
  \leq 2 \, k_i \,
  \min\big\{|s_2-s_1|,|I|\big\}
  \,,
$$
where
$$
  k_i :=
  \begin{cases}
    \|\kappa_1\|_\infty
    & \mbox{if} \quad  i=1 \,,
    \\
    \|\kappa_1\|_\infty + \|\kappa_2\|_\infty
    & \mbox{if} \quad  i=2 \,,
    \\
    \|\kappa_2\|_\infty
    & \mbox{if} \quad  i=3 \,.
  \end{cases}
$$
\end{Lemma}
\begin{proof}
It follows from the Serret-Frenet equations~(\ref{Frenet})
and~(\ref{curvatures}) that
\begin{equation*}
  \big|e_i(s_2)-e_i(s_1)\big|
  \leq 2 \left| \int_{s_1}^{s_2} |\dot{e}_i| \right|
  \leq 2 \, k_i
  \left| \int_{s_1}^{s_2} \chi_I \right|
  ,
\end{equation*}
which immediately establishes the assertion.
\end{proof}

As a consequence of Lemma~\ref{Lem.closing},
we get the inequality
\begin{equation}\label{closing}
  e_i(s_2) \cdot e_i(s_1)
  \ \geq \ 1-2 \, |I|^2 \, k_i^2
  \,, \qquad
  i\in\{1,2,3\}
  \,.
\end{equation}
In particular,
since~$e_1$ is the tangent vector of~$\Gamma$,
we obtain that the curve is not self-intersecting
provided
$
  |I| \, \|\kappa_1\|_\infty < 1
$.
A stronger sufficient condition ensures the injectivity of~$\tube$:
\begin{Proposition}\label{Prop.injectivity}
Let~$\Gamma$ be determined
by the curvature functions~(\ref{curvatures}).
Then the hypotheses~(\ref{Ass.basic}) hold true provided
$$
  \max\left\{
  4 \, |I|^2 \, \|\kappa_1\|_\infty^2
  \,, \
  4 \, a \, \big(\|\kappa_1\|_\infty+\|\kappa_2\|_\infty\big)
  \right\} \ < \ 1
  \,.
$$
\end{Proposition}
\begin{proof}
The idea is to observe that it is enough to show
that the mapping~$\Gamma_t$ from~$\Real$ to~$\Real^3$
defined by
$$
  \Gamma_t(s)
  := \Gamma(s) + t_\mu \, \rot_{\mu\nu}(s) \, e_\nu(s)
$$
is injective for
any \emph{fixed} $t \in \Real^{2}$ such that $|t|<a$
and \emph{arbitrary} matrix-valued function
$
  (\rot_{\mu\nu}) : \Real \to \mathsf{SO}(2)
$.
Let us assume that there exist $s_1 < s_2$
such that $\Gamma_t(s_1)=\Gamma_t(s_2)$.
Then
\begin{align*}
  0 \ = \ & \Gamma(s_2)-\Gamma(s_1)
  \\
  & + t_\mu \left\{
  \big[\rot_{\mu\nu}(s_2)-\rot_{\mu\nu}(s_1)\big] \;\! e_\nu(s_1)
  + \rot_{\mu\nu}(s_2) \;\! \big[e_\nu(s_2)-e_\nu(s_1)\big]
  \right\}
  \,.
\end{align*}
Taking the inner product of both sides of the vector identity
with the tangent vector $e_1(s_1)$ and writing
the difference $\Gamma(s_2)-\Gamma(s_1)$ as an integral,
we arrive at the following scalar identity
\begin{equation*}
  0 = \int_{s_1}^{s_2} e_1(s_1) \cdot e_1(\xi) \, d\xi
  + t_\mu \, \rot_{\mu\nu}(s_2) \;\! \big[e_\nu(s_2)-e_\nu(s_1)\big]
  \cdot e_1(s_1)
  \,.
\end{equation*}
Applying Lemma~\ref{Lem.closing} together with
the first inequality of~(\ref{closing}),
recalling the orthogonality of~$(\rot_{\mu\nu})$
and using obvious estimates,
we obtain
\begin{equation*}
  0 \geq (s_2-s_1) \, \left(
  1-2 \, |I|^2 \, k_1^2
  - 2 \, a \, k_2
  \right)
  .
\end{equation*}
This provides a contradiction for curves satisfying
the inequality of Proposition, unless $s_1=s_2$.
\end{proof}
\begin{Remark}
The ideas of this Appendix do not restrict to the special class
of tubes about curves determined by~(\ref{curvatures}).
Indeed, assuming only the existence of an appropriate Frenet frame
for the reference curve (\cf~\cite{ChDFK}),
more general sufficient conditions,
involving integrals of curvatures,
could be derived.
\end{Remark}
%

\section*{Acknowledgment}

The authors are grateful to Timo Weidl for pointing out the
presented problem to them.
The work has partially been supported by
the Czech Academy of Sciences and its Grant Agency
within the projects IRP AV0Z10480505 and A100480501,
and by DAAD within the project D-CZ~5/05-06.
T.E. has partially been supported by the ESF European programme SPECT and
D.K. has partially been supported by FCT/\-POCTI/\-FEDER, Portugal.

%
%
\providecommand{\bysame}{\leavevmode\hbox to3em{\hrulefill}\thinspace}
\providecommand{\MR}{\relax\ifhmode\unskip\space\fi MR }
\providecommand{\MRhref}[2]{%
  \href{http://www.ams.org/mathscinet-getitem?mr=#1}{#2}
}
\providecommand{\href}[2]{#2}
\end{document}